\documentclass[12pt]{iopart}

\usepackage{comment}
\usepackage{graphicx}
\usepackage{float}
\usepackage{subfigure}
\usepackage{tabularx}
\usepackage{cite}

\usepackage{caption}
\usepackage{amsmath,amsfonts,amssymb}
\usepackage{iopams}
\usepackage{subfiles} % Best loaded last in the preamble
\usepackage{url}
\usepackage{color}
\usepackage{xcolor}
\usepackage{bigints}

\usepackage{tikz}
\usepackage[hidelinks]{hyperref}
\usepackage{orcidlink}

\newcommand{\lambdabar}{{\mkern0.75mu\mathchar '26\mkern -9.75mu\lambda}}

\begin{document}

\title[QScatter: Numerical Framework for Fast Prediction of Particle Distributions in Electron-Laser Scattering]{QScatter: Numerical Framework for Fast Prediction of Particle Distributions in Electron-Laser Scattering}

\author{
    Óscar Amaro \orcidlink{0000-0003-0615-0686},
    Marija Vranic \orcidlink{0000-0003-3764-0645} 
}

\address{
GoLP/Instituto de Plasmas e Fus\~{a}o Nuclear, Instituto Superior T\'{e}cnico, Universidade de Lisboa, 1049-001 Lisbon, Portugal
}
\ead{\footnotesize
\href{mailto:oscar.amaro@tecnico.ulisboa.pt}{oscar.amaro@tecnico.ulisboa.pt}
, \href{mailto:marija.vranic@tecnico.ulisboa.pt}{marija.vranic@tecnico.ulisboa.pt}
}\normalsize

\vspace{10pt}

\begin{abstract}

The new generation of multi-PetaWatt laser facilities will allow tests of Strong Field QED, as well as provide an opportunity for novel photon and lepton sources. The first experiments are planned to study the (nearly) head-on scattering of intense, focused laser pulses with either relativistic electron beams or high-energy photon sources. 
In this work, we present a numerical framework that can provide fast predictions of the asymptotic particle and photon distributions after the scattering. The works presented in this manuscript includes multiple features such as spatial and temporal misalignment between the laser and the scattering beam, broadband electron beams, and beam divergence. The expected mean energy, energy spread, divergence or other observables are calculated by combining an analytical description and numerical integration. This method can provide results within minutes on a personal computer, which would otherwise require full-scale 3D QED-PIC simulations using thousands of cores. The model, which has been compiled into an open-source code \texttt{QScatter}, may be used to support the analysis of large-size data sets from high-repetition rate experiments, leveraging its speed for optimization or reconstruction of experimental parameters.
    
\end{abstract}

\section{Introduction}\label{sc:introduction}

Strong-field quantum electrodynamics (SFQED) is a rapidly developing research field that studies the interaction between matter and intense electromagnetic fields. In recent years, there has been a growing interest in this area due to the availability of high-intensity laser sources, which enable the exploration of novel physical phenomena, with experiments being planned for the near future: 
% facilities
ELI \cite{ELI}, Apollon \cite{APOLLON}, CoReLS \cite{CoReLS}, FACET-II \cite{FACET-II,meurenSeminalHEDPResearch2020}, LUXE \cite{abramowiczLetterIntentLUXE2019, abramowiczConceptualDesignReport2021}, EXCELS \cite{EXCELS}, ZEUS\cite{ZEUS}, Omega Laser Facility \cite{OMEGA}, HIBEF \cite{HIBEF}, among others. 
The proposed experimental setups consist of scattering of intense, focused laser pulses with either relativistic electron beams or high-energy photons, allowing for precision studies of radiation reaction (the recoil on the charged particles that emit high-energy photons) and electron-positron production in the lab.

In these studies, different regimes of radiation reaction can be identified based on the energies of the probe particles and the strength of the electromagnetic fields involved. 

Recent experiments have demonstrated electron energy loss that can be attributed to radiation reaction \cite{coleExperimentalEvidenceRadiation2018, poderExperimentalSignaturesQuantum2018}.
For low enough electron energies and laser intensities, the effect of radiation on the radiation emitting electrons can be described through a continuous correction to the equation of motion, e.g. the Landau-Lifshitz equation  \cite{landauClassicalTheoryFields1980}. 
For intermediate values of energies and intensities, the evolution of the particle distribution can be modeled by a Fokker-Planck equation, where the energy of each emitted photon is assumed to be much smaller than the emitting lepton  \cite{neitzStochasticityEffectsQuantum2013, vranicQuantumRadiationReaction2016}.
In this case, the particle can still lose a significant fraction of its energy through multiple emissions of low-energy photons. 
For higher energies and intensities, a single photon emission can lead to almost complete energy depletion of the parent lepton, and thus a Boltzmann/transport equation becomes a better description. A comparison between the different models of radiation reaction and their range of validity can be found in the recent literature \cite{ildertonRadiationReactionStrong2013,vranicClassicalRadiationReaction2016,nielQuantumClassicalModeling2018a}.

Besides radiation reaction, a phenomena likely to occur in electron-laser collisions is Breit-Wheeler \cite{breitCollisionTwoLight1934} electron-positron pair production. 
The pair production rates for this process are known for particles in a constant intense background field, and can be mapped onto a plane wave scenario. 
There are also extended models that take into account the fact that the laser is a wavepacket with a temporal envelope. 
% Blackburn 2017
For example, the authors in Ref. \cite{blackburnScalingLawsPositron2017} derive approximate scaling laws for the expected positron yield for both photon-laser and electron-laser collisions.
Further theoretical work was pursued
% Mercuri-Baron 2021
by \cite{mercuri-baronImpactLaserSpatiotemporal2021}, where the authors investigated photon-laser scattering for different Laguerre-Gauss modes of the laser, and in
% Amaro 2021
\cite{amaroOptimalLaserFocusing2021} where optimal focusing conditions were found for maximizing the number of pairs from the scattering between Gaussian laser pulses and different electron beam profiles.

% non-perturbative
Since the strong-field regime of QED is characterized by the breakdown of perturbative approaches, it requires the development of new theoretical and computational tools. Numerical modeling of experiments with enough spatiotemporal resolution and accounting for all relevant scattering parameters becomes too computationally expensive if relying on full-scale 3D PIC and Monte Carlo simulations, especially if the analysis requires multiple parameter studies. New methods are currently being evaluated for accelerating the evaluation of QED rates (e.g. Machine-Learning based \cite{badialiMachinelearningbasedModelsParticleincell2022}, and Chebyshev polynomial fits  \cite{montefioriSFQEDtoolkitHighperformanceLibrary2023}).

\begin{figure}[H]
    \centering
    \includegraphics[width=300pt]{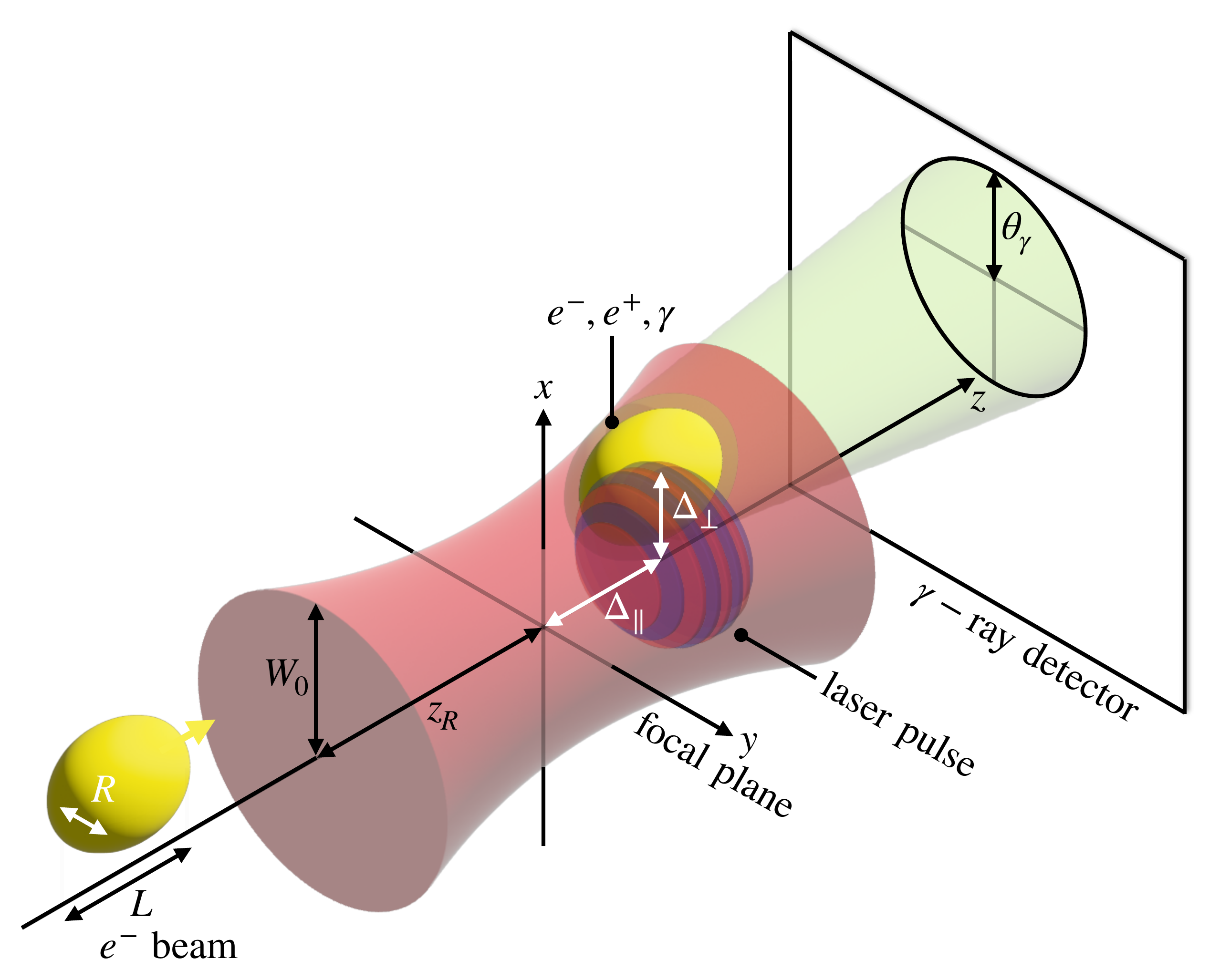}
    \caption{Scattering of a relativistic electron beam (yellow) and a counter-propagating laser pulse (red and blue). Electron-positron pairs and high energy photons can be produced in the interaction, and their number, spectra and divergence will depend on the precise geometry of the collision.} 
    \label{fig:scatt}
\end{figure}

% this paper
In this paper, we describe the development of a reduced semi-analytical model for the final energy and angular distribution of particles in electron-laser scattering (see figure \ref{fig:scatt}).
% this paper's goal
This approach simplifies the calculation for various beam geometries, focusing and collision synchronization.  
% computational cost 
The results can be obtained within minutes on a personal computer. 
% OSIRIS
We perform benchmarks with the fully-relativistic particle-in-cell code OSIRIS \cite{fonsecaOSIRISThreeDimensionalFully2002}, and demonstrate that with this model quantitative predictions can be obtained without the need for the full-scale simulations. This can be particularly useful for real-time parameter scans during the course of an experiment.

% Structure of paper
%
This manuscript is structured as follows. In section \ref{sc:a0eff}, we discuss the mapping between plane wave (PW) models and more realistic ``3D'' scattering setups.
In section \ref{sc:spectra}, we use this approach to derive the final spectra of photons and electrons after scattering with a finite, focused laser pulse. 
In section \ref{sc:photonSpectra}, we apply an approximate semi-classical numerical model for the photon energy distribution.
In section \ref{sc:angular},  we investigate the electron and photon angular distribution as well as positron yield. 
Finally, the conclusions are presented in section \ref{sc:conclusions}.

% dNda
\section{Effective intensity in focused laser scattering}\label{sc:a0eff}

This section introduces the theoretical framework for describing the distribution of particles in the maximum effective intensity of the laser they interact with along their ballistic trajectories. This approach aims at simplifying the mapping between 1D plane wave models to more realistic 3D-focused-laser environments.

To better understand the derivation of the effective intensity distribution, it is useful to first recall the case of a plane wave pulse. When a beam of particles collides with this laser pulse (there is no focusing, the wavepacket has a temporal envelope and no transverse structure), all particles get to interact with the maximum intensity. For this case, the distribution of particles $\mathrm{d}N_b$ according to the peak laser vector potential they interact with $a_{0,\mathrm{eff}}$ is of the form $\mathrm{d}N_b/\mathrm{d}a_{0,\mathrm{eff}} \sim \delta(a_{0,\mathrm{eff}}-a_0)$, as all particles experience the maximum vector potential $a_0$.

However, in the case of a diffraction-limited focused laser, not all beam particles (which can be either ultra-relativistic electrons or photons)  interact with the peak laser intensity, but rather with a maximum $a_{0,\mathrm{eff}}$ that depends on the longitudinal and transverse offsets from the focus at the collision time. If one considers a  cylindrical coordinate system $\vec{r}=(\rho, \phi, z)$, each beam particle is assigned an effective vector potential $a_{0,\mathrm{eff}}(\rho, \phi, z)$ according to the distribution function

\begin{equation}
    \frac{\mathrm{d}N_b}{\mathrm{d}a_{0,\mathrm{eff}}}(a_{0,\mathrm{eff}}) = \int_V 2~\delta(a(\vec{r})-a_{0,\mathrm{eff}}) ~n_b(\vec{r}) ~\mathrm{d}V =  \int_S \frac{2~n_b \ \mathrm{d}S}{||\nabla a||}.
    \label{eq:dNda0eff}
\end{equation}
where $a_{0,\mathrm{eff}}$ is the argument of the distribution function, $a(\vec{r})$ is the effective vector potential as a function of the coordinates (to be specified later), $n_b$ is the particle number density in the lab frame, the factor 2 accounts for crossing time (at twice the speed of light for ultrarelativistic particles), and the surface integral (right-hand side) is evaluated at $a=a_{0,\mathrm{eff}}$, . The functional expression above assigns each macro-particle (which is equivalent to a volume element in the electron distribution) a spatial coordinate corresponding to the position along its trajectory where it interacts with the maximum laser field.
The volumetric integral then assigns the volume element to the correct bin in the $a_{0,\mathrm{eff}}$ distribution. 
As a Gaussian focusing geometry is cylindrically symmetric around the laser propagation axis, we can convert equation \eqref{eq:dNda0eff} to a surface integral, where $\mathrm{d}S = \rho \sqrt{\mathrm{d}\rho^2+\mathrm{d}z^2} ~\mathrm{d}\varphi = \rho \sqrt{1+(\partial\rho/\partial z)^2} ~\mathrm{d}z ~\mathrm{d}\varphi$ is calculated at the isosurface that is by definition perpendicular to the gradient of the vector potential given by $||\nabla a||= \sqrt{\left(\partial a/\partial \rho\right)^2+\left(\partial a/\partial z\right)^2}$.

The distribution becomes a function of both the beam density profile and the laser's spatial structure (neglecting the wave's phase, and in particular its wavefront curvature). For example, the ideal Gaussian laser has a spatial dependence
\begin{equation}
    a (\rho, \phi, z) = \frac{a_0}{\sqrt{1+(z/z_R)^2}} \exp \left( -\frac{\rho^2/W_0^2}{1+(z/z_R)^2} \right)
    \label{eq:a0eff}
\end{equation}
\noindent where $W_0$ is the spotsize, $z_R\equiv \pi W_0^2/\lambda$ the Rayleigh length, and $\lambda$ is the laser central wavelength.

\begin{figure}[H]
    \centering
    \includegraphics[width=440pt]{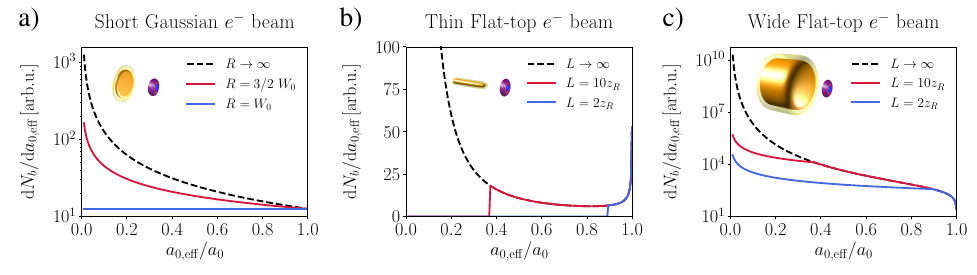}
    \caption{Particle distributions for different geometries, where $\lambda=1~\mu \mathrm{m}$ and $W_0=2~\mu \mathrm{m}$. The dashed line corresponds to the limit of either infinite beam radius (Short beam) or infinite length (Thin and Wide beams).}
    \label{fig:geometries}
\end{figure}

% beam geometry
For an arbitrary beam density profile $n_b(\vec{r})$ with length $L$ and radius $R$, it can be challenging to analytically compute the particle distribution in $a_{0,\mathrm{eff}}$. However, in the limiting cases of a Short beam ($L \ll z_R$, transverse Gaussian density), Wide beam ($R \gg W_0$, longitudinal and transverse flat-top density) and Thin beam ($R \ll W_0$, longitudinal flat-top density) these distributions can be calculated. In figure \ref{fig:geometries}, we present several instances of these geometries and their respective distributions.
In table \ref{tb:dist}, we collect the distributions for the three above-mentioned geometries, generalized to the case of non-synchronized scattering, i.e., when the particle beam and laser collide outside of the focus.

% Distributions are not equally normalized to facilitate comparison.

\begin{table}
{\setlength{\extrarowheight}{10pt}
\begin{tabularx}{\textwidth}{l l} 
\hline
Setup & Particle distribution for temporally unsynchronized beams \\[5pt]
\hline\hline \\[-15pt]
$\begin{array}{cc}
    \textrm{Wide}\\[-10pt] \textrm{beam}
\end{array}$ & \footnotesize$ \dfrac{\mathrm{d}N_b}{\mathrm{d}a} = \dfrac{2\pi ~n_b ~W_0^2 ~z_R}{ a} \left( \dfrac{z_+}{z_R} \left(1+\dfrac{1}{3} \left( \dfrac{z_+}{z_R} \right)^2 \right) \theta(a_{z_+}-a) -\dfrac{z_-}{z_R} \left(1+\dfrac{1}{3} \left( \dfrac{z_-}{z_R} \right)^2 \right) \theta(a_{z_-}-a) \right.$\\[25pt]
 & \footnotesize$\left. ~~~~~~~~~~~~+\dfrac{\sqrt{a_0^2-a^2}}{3~a} \left( 2+\dfrac{a_0^2}{a^2}\right) \left(\theta(a-a_{z_+})\pm\theta(a-a_{z_-})~ \right) \right) $\\[25pt]
 $\begin{array}{cc}
    \textrm{Thin} \\[-10pt] \textrm{beam}
\end{array}$ & \footnotesize$\dfrac{\mathrm{d}N_b}{\mathrm{d}a} = 
        \dfrac{2N_b z_R}{L}\dfrac{a_0^2}{a^2} \dfrac{1}{\sqrt{a_0^2-a^2}} \left(\theta(a-a_{z_+})\pm\theta(a-a_{z_-})~\right)
        $ \\[35pt]
 $\begin{array}{cc}
    \textrm{Short}\\[-10pt] \textrm{beam}
\end{array}$ &
\footnotesize $\dfrac{\mathrm{d}N_b}{\mathrm{d}a} = N_b \dfrac{W^2}{R^2} \dfrac{1}{a} \left(\dfrac{a}{a_0}~\dfrac{W}{W_0} \right)^{W^2/R^2} 
    I_0\left(2\dfrac{\Delta_\bot}{R} \dfrac{W}{R} \log^{1/2}\left(\dfrac{a_0}{a}~\dfrac{W_0}{W}\right) \right) \exp\left(-\dfrac{\Delta_\bot^2}{R^2}\right) \theta(a-a_\parallel)$ \\[25pt]
 \hline
 \end{tabularx}
 }
\caption{Particle distributions for arbitrary temporal synchronization. A shorter notation is used $a \equiv a_{0,\mathrm{eff}}<a_0$. Here, $a_{z} \equiv a_0/\sqrt{1+(L/4z_R)^2}$ is the $a_{0,\mathrm{eff}}$ associated with the integration limits imposed by the longitudinal size of the electron beam, $N_b$ represents the total number of particles in the beam, $n_b$ is the beam density, $R$ and $L$ are the beam radius and length respectively. The laser spot size is $W_0$ in the focal plane and $W=W_0\sqrt{1+\Delta_\parallel^2/z_R^2}$ outside, $z_R\equiv \pi W_0^2/\lambda$ is the Rayleigh length, and $\Delta_\parallel$, $\Delta_\bot$ are the displacements of the collision center from the laser propagation axis and focal plane, respectively. The $\pm$ sign corresponds to situations where $\Delta_\parallel<L/4$ or $\Delta_\parallel>L/4$ respectively, $\theta(x)$ is the Heaviside Theta function, $z_\pm=\Delta_\parallel \pm L/4$, $a_{z_\pm} = a_0/\sqrt{1+(z_\pm/z_R)^2}$ and $a_\parallel = a_0/\sqrt{1+(\Delta_\parallel/z_R)^2}$.
%, 
}
\label{tb:dist} 
\end{table}

% our work
We first introduced the concept of the $\mathrm{d}N_b/\mathrm{d}a_{0,\mathrm{eff}}$ distribution in our previous work \cite{amaroOptimalLaserFocusing2021}, where we applied it to the optimization of the number of pairs produced in electron-laser scattering. 
We accomplished that by generalizing the scaling law previously derived for positron production in a plane wave  \cite{blackburnScalingLawsPositron2017}. 
This approach is cost-effective and amenable to analytical calculations pairing with PW models for other observables. 
In the next section, we apply our method to estimate the final electron energy spectrum in a focused laser scattering.

\section{Electron and photon spectra}\label{sc:spectra}

In this section, we apply the effective interaction intensity distributions to derive the asymptotic energy distributions of the electrons after the scattering. Simulation parameters for the given examples can be found in appendix \ref{app:A}.

\subsection{Classical Radiation Reaction}\label{ssc:crr}

If the scattering between the laser and the electron beam is in the classical regime of interaction, one can assign a deterministic trajectory to an electron within the beam. 
If there is considerable radiation emission (e.g. a few percent of the electron energy is radiated) during the interaction, then we can describe the motion though the Landau-Lifshitz equation, which has an exact solution in a plane wave \cite{piazzaExactSolutionLandauLifshitz2008}. This solution calculates the full space-time trajectory of the particle as it interacts with the laser field (in configuration and momentum space). For the case of a pulse with a temporal envelope, this solution can be condensed into a scaling law for the final energy of the electron \cite{vranicAllOpticalRadiationReaction2014}

\begin{equation}
    \gamma_{f}=\frac{\gamma_{0}}{1+c_{RR}~ \gamma_{0}~a_0^2}~, \quad c_{RR} \equiv (1-\cos \theta)^{2} \frac{\eta}{3} \frac{e^{2} \omega_{0}^{2}}{m c^{3}} \tau_{0}.
    \label{eq:PWCRR}
\end{equation}

\noindent Here, $\gamma_0$ is the initial electron energy, $\tau_0$ is the laser pulse duration, $\omega_0$ the fundamental laser frequency, 
$c_{RR}$ is a numerical factor quantifying radiation reaction that depends on the laser shape and scattering angle, $\eta$ is a numerical factor associated with the temporal shape of the laser pulse, $\theta$ is the interaction angle between the laser wavefront and the colliding electron, $e$ is the elementary electric charge, $m$ the electron mass, and $c$ is the speed of light in vacuum.

\begin{figure}[H]
    \centering
    \includegraphics[width=440pt]{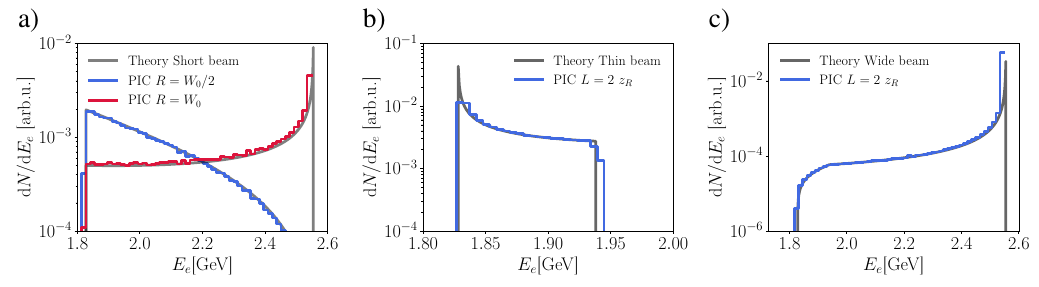}
    \caption{\raggedright Final electron energy distributions for Short, Thin and Wide beam geometries after classical radiation reaction. Parameters: $\gamma_0=5000$, $a_0=12$, $\lambda=0.8~\mu \mathrm{m}$, $\tau_0=50~\omega_0^{-1}$. The exact formulas for the theoretical distribution are presented in Table \ref{tb:RRspectra}}
    \label{fig:CRR3Dspectra}
\end{figure}

Equation \eqref{eq:PWCRR} shows a 1-to-1 relation between the final energy and the $a_0$ of the plane wave (fixing all other parameters), which means it is possible to directly calculate the final energy for each particle, provided we know the maximum laser intensity and $\gamma_0$. Combining the effective intensity distributions from table \ref{tb:dist} and the equation \eqref{eq:PWCRR}, we obtain the full energy distribution function after the interaction

\begin{equation}
    \dfrac{\mathrm{d}N_b}{\mathrm{d}\gamma_f}(\gamma_f) = \dfrac{\mathrm{d}N_b}{\mathrm{d}a_{0,\mathrm{eff}}}  ~ \left( \dfrac{\mathrm{d}\gamma_f}{\mathrm{d}a_{0,\mathrm{eff}}} \right)^{-1}
    \label{eq:dNda2dNdgf}
\end{equation}

where the second part is the inverse of the derivative of equation \ref{eq:PWCRR} with $a_0 = a_{0,\mathrm{eff}}$. In figure \ref{fig:CRR3Dspectra}, we compare the distributions obtained from the analytical expression \eqref{eq:dNda2dNdgf} against PIC simulations. Here, electrons radiate energy following the Landau-Lifschitz classical radiation reaction model, while the quantum radiation reaction/QED module was turned off in the simulation parameters to demonstrate the applicability of the model. For the different geometries and aspect ratios $R/W_0$, we find good agreement between the predicted distributions and simulation results. It is worth noting that the cut-offs in the spectra (specially for the Short and Thin beams) is a feature of deterministic radiation reaction; however, this will be smoothed out if the initial electron beam has large enough energy spread or a stochastic quantum radiation reaction model is used.

The summary of formulas for the final energy distributions of electrons is given in table \ref{tb:RRspectra}. We show the expression for the same examples as in the previous table, where for the sake of simplicity we consider the scattering synchronized and aligned  ($\Delta_\parallel =0, \Delta_\bot =0$). The distributions can be derived for any case (from table \ref{tb:dist} or not) following the procedure that we just outlined.

% table of energy distributions
\begin{table}[H]
{\setlength{\extrarowheight}{10pt}
\begin{tabularx}{0.95\textwidth}{l l} 
\hline
Setup & Energy distributions after classical radiation reaction \\[5pt]
\hline \\[-15pt]
    Wide beam & $\dfrac{\mathrm{d}N_b}{\mathrm{d}\gamma} = 4\pi ~n_b ~W_0^2 ~z_R \dfrac{\gamma_0/\gamma}{\gamma_0-\gamma} ~\left( \dfrac{L(L^2+16z_R^2)}{128z_R^3}~ \theta_1 \right. $\\[15pt]
    & $\left. \quad\quad +\dfrac{1}{6(\gamma_0-\gamma)^{3/2}} (c_{RR} \gamma \gamma_0 a_0^2+2(\gamma_0-\gamma)) \sqrt{c_{RR} \gamma \gamma_0 a_0^2+\gamma-\gamma_0} ~(1-\theta_1) \right)$ \\[15pt]
    Thin beam & $\dfrac{\mathrm{d}N_b}{\mathrm{d}\gamma} = \dfrac{2N_b ~z_R}{L} \dfrac{\gamma_0}{(\gamma_0-\gamma)^{3/2}} \dfrac{c_{RR} \gamma_0 a_0^2}{\sqrt{c_{RR} \gamma \gamma_0 a_0^2+\gamma-\gamma_0}} ~\theta_1 ~\theta_2$ \\[25pt] 
    Short beam & $\dfrac{\mathrm{d}N_b}{\mathrm{d}\gamma} = N_b ~\dfrac{W_0^2}{2R^2} \dfrac{\gamma_0/\gamma}{\gamma_0-\gamma} \left( \dfrac{\gamma_0-\gamma}{c_{RR} \gamma_0 \gamma a_0^2} \right)^{\dfrac{W_0^2}{2R^2}} ~ \theta_2$ \\[15pt] 
 \hline
 \end{tabularx}
 }
\caption{\raggedright Electron spectra after classical scattering with focused laser pulses for different beam geometries. Distributions are obtained from table \ref{tb:dist} and the scaling law \eqref{eq:PWCRR}. The minimum electron energy is $\gamma_\mathrm{min} \equiv \gamma_0/(1+c_{RR} \gamma_0 a_0^2)$ and the energy at branch transition is $\gamma_z \equiv \gamma_0/(1+c_{RR} \gamma_0 a_z^2)$. $\theta_1\equiv \theta(\gamma-\gamma_z)$, $\theta_2 \equiv \theta(\gamma-\gamma_\mathrm{min})$}
 \label{tb:RRspectra} 
\end{table}

\subsection{Quantum Radiation Reaction and misaligned beams: semi-analytical approach}\label{ssc:qrr}

In the previous sub-section, we considered the case of deterministic, classical radiation reaction. As the laser field and the electron energy increase, the photon emission becomes stochastic due to its quantum nature and tends to spread the distribution function of the photon emitting electrons. 
There is no general analytical solution for the evolution of the coupled electron and photon distributions, even in the simplest field configurations. 
However, the plane wave case is easy to simulate numerically in one dimension. 
In this subsection, we show that we can use specific plane wave solutions from 1D simulations to reconstruct a more general 3D result.
Similarly to the deterministic case, the energy lost to photons is expected to increase as the laser $a_0$ increases and the ratio $R/W_0$ decreases.

\begin{figure}[H]
    \centering
    \includegraphics[width=440pt]{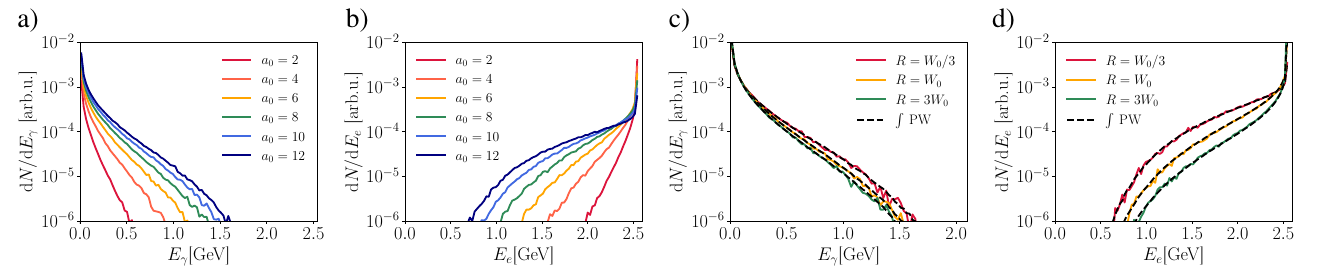}
    \caption{\raggedright Reconstructing particle spectra in focused laser collisions using pulsed plane wave datasets. a) and b) photon and electron final spectra in pulsed plane wave collisions, respectively. c) and d) Final spectra from a 3D simulation of the interaction of a diffraction-limited (Gaussian) laser pulse against the reconstruction from 1D plane wave samples.}
    \label{fig:QRR1Dto3D}
\end{figure}

In figures \ref{fig:QRR1Dto3D} a) and b) we show simulation results for the final photon and electron spectra after scattering with a pulsed plane wave. Analogously with the classical case, there are more electrons in the low energy tail of the distribution for the high values of $a_0$. 
To map these results to 3D, similarly to equation \eqref{eq:dNda2dNdgf}, we weight the plane wave spectra $dN^{PW}/d\gamma$ by the $a_{0,\mathrm{eff}}$ distribution function

\begin{equation}
    \dfrac{\mathrm{d}N}{\mathrm{d}\gamma}^{3D}(\gamma; a_0) 
    \sim \sum_{a_{0,\mathrm{eff}}} \dfrac{\mathrm{d}N}{\mathrm{d}\gamma}^{PW}(\gamma, a_{0,\mathrm{eff}})~\dfrac{\mathrm{d}N_b}{\mathrm{d} a_{0,\mathrm{eff}}}(a_{0,\mathrm{eff}})
\end{equation}

\noindent where $a_{0,\mathrm{eff}}$ is a uniformly distributed, discrete set of $a_0$ values with enough resolution to represent the full effective intensity distribution.

By running several PW simulations in parallel, this approach can be used to calculate the asymptotic distributions expected after the interaction with a focused Gaussian laser pulse (including and all the variations on alignment), which has orders of magnitude lower computational cost than one full-scale 3D or quasi-3D simulation. In figures \ref{fig:QRR1Dto3D} c) and d) we present the results of quasi-3D simulations in color against the reconstruction method presented before, showing good agreement. We used PW simulations with $a_0$ from $0$ to $12$ in step increments of $0.2$.

\begin{figure}[H]
    \centering
    \includegraphics[width=440pt]{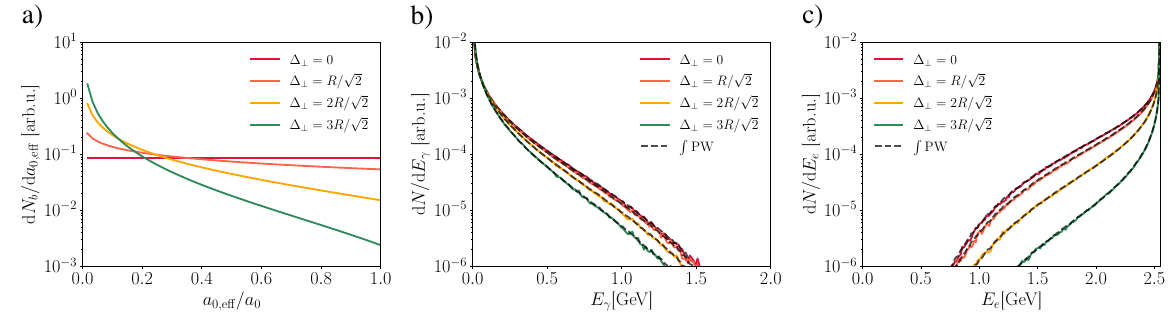}
    \caption{\raggedright Perpendicular displacement of Short beam. Left: $a_{0,\mathrm{eff}}$ distribution. Middle: photon spectra. Right: electron spectra.  Same main parameters as in figure \ref{fig:QRR1Dto3D}, with maximum $a_0=12$.}
    \label{fig:QRRDprp}
\end{figure}

The same PW dataset can be effectively used to reconstruct the electron and photon distributions from a collision with a transverse offset (presented in figure \ref{fig:QRRDprp}).

\begin{figure}[H]
    \centering
    \includegraphics[width=440pt]{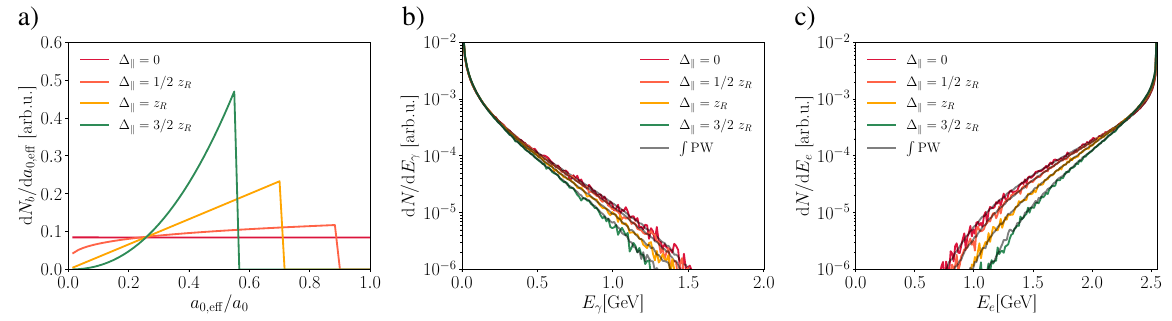}
    \caption{\raggedright Parallel displacement of Short beam. Left: $a_{0,\mathrm{eff}}$ distribution. Middle: photon spectra. Right: electron spectra.  Same main parameters as in figure \ref{fig:QRR1Dto3D}, with maximum $a_0=12$.}
    \label{fig:QRRDpll}
\end{figure}

It is also possible to reconstruct the expected distributions for longitudinal (temporal) offsets (see figure \ref{fig:QRRDpll}).

In later sections, this approach is applied to other observables from laser-beam scattering.

\section{The approximate photon spectra}\label{sc:photonSpectra}

In this section, we discuss the possibility of computing an approximate asymptotic photon spectrum without resorting to PIC simulations. This can replace the 1D plane wave calculations of the photon spectra from the previous section. 

As an electron interacts with a wavepacket, it loses energy, on average, following the semi-classical equation of motion (the Lorentz force is omitted in the equation, we just display the radiation reaction term)

\begin{equation}
    \left( \dfrac{\mathrm{d}p}{\mathrm{d}t} \right)_{RR} = -~g(\chi_e)~P_{cl}/c, ~P_{cl} = \dfrac{2~\alpha ~c}{3 ~\lambdabar_c} m_e c^2 \chi_e^2
    \label{eq:dpdtRR}
\end{equation}

\noindent where $\chi_e \sim 2 ~\gamma~a/a_S$ is the lepton quantum nonlinearity parameter; $a_S$ is the normalized Schwinger vector potential; $P_{cl}$ the classical synchrotron-like emitted power, $\alpha$ is the fine-structure constant; $\lambdabar_c$ is the Compton wavelength, and $g(\chi_e)$ the electron Gaunt-factor (which corrects the average emitted power to be consistent with a full quantum approach).

\begin{figure}[H]
    \centering
        \includegraphics[width=340pt]{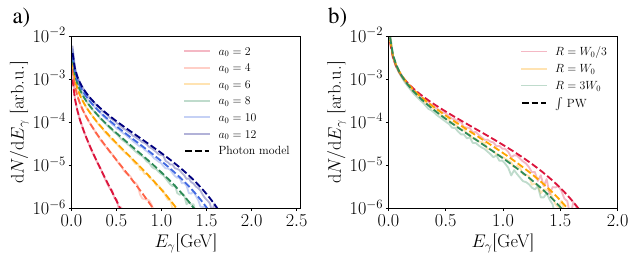}
        \caption{Left: comparison between final cumulative photon spectra in electron-laser PW interaction for varying $a_0$ in simulations (solid) and reconstructed (dashed). Right: comparison between final cumulative photon spectra in electron-laser PW interaction for varying $R/W_0$ in simulations (solid) and reconstructed (dashed).}
    \label{fig:diffrate}
\end{figure}

According to equation \eqref{eq:dpdtRR},  for each phase of the laser $\phi$ we can assign an associated average electron energy $\gamma(\phi)$ and combine this with an instantaneous laser field vector potential $a(\phi)$. For each of these pairs of values, we compute the quantum synchrotron spectrum (in local constant field approximation) and integrate them to obtain the final, cumulative photon spectrum as shown in figure \ref{fig:diffrate}.

This results in a very good estimate of the photon spectra. Figure \ref{fig:diffrate} also shows small deviations between the simulations and the theoretically predicted reconstructions. These deviations can be attributed to the neglected energy spread of the photon emitting electrons. This model could be further improved by applying a Boltzmann equation solver to capture a self-consistent, multi-particle evolution of all the energy momenta.

\section{Angular distribution of electrons,  photons and positron yield}\label{sc:angular}

After having addressed the average energy, energy spread and the evolution of the electron energy distribution function during the laser-electron scattering, in this section we discuss the possibility of incorporating several scaling laws for different scattering observables into the model. In particular, the highlight is given to the angular distribution and of both electrons and photons, as this can be directly measured in experiments. For completeness, we also mention how the same method is applied to calculate the positron yield, which was addressed in more detail in our previous work \cite{amaroOptimalLaserFocusing2021}. 

\subsection{Photon angular distribution}

% intro
Relativistic electrons emit photons within an angle $\sim 1/\gamma$ around their propagation direction. 
When an electron interacts head-on with a linearly polarized laser (LP), the transverse momentum is of the order of the laser vector potential, so the expected opening angle of the emission is $\sim \left<a\right>/\gamma$ in the  polarization direction.
The third direction, perpendicular both to the propagation and the polarisation still has an expected typical divergence of $\sim 1/\gamma$. Here $a$ and $\gamma$ refer to the instantaneous laser amplitude and electron energy. To account for a laser temporal envelope (in particular, we use a $\sin^2$ pulse envelope) and linear polarization, the average instantaneous value of the laser vector potential is twice decreased by a factor of $\sqrt{2}$ compared to the peak field. This gives an overall expected root-mean-squared photon angle of $\sim 0.5~a_0/\gamma_0$.

\begin{figure}[H]
    \centering
    \includegraphics[width=450pt]{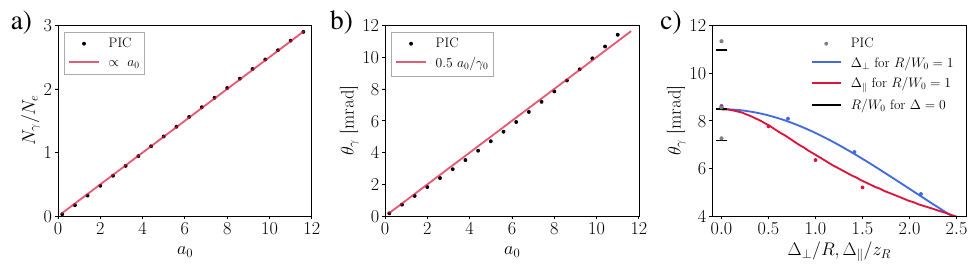}
        \caption{Left: number of photons per electron (only including photons above 1 MeV energy). Middle: photon angle rms in plane wave. Right: photon angle rms for setups using focused lasers.}
    \label{fig:thetaPhotons}
\end{figure}

For moderate to high laser intensity, the number of photons above a certain energy scales linearly with $a_0$ (see figure \ref{fig:thetaPhotons} a). 
Since the number of photons emitted by the electrons is a function of $a_0$, this has to be included in the weighted average over the instantaneous emission angle to obtain the expected final divergence of the signal.

As shown in figure \ref{fig:thetaPhotons}, using this information and pairing it with the effective intensity distributions allows predicting the final photon divergence for diffraction-limited laser-electron collisions, including spatio-temporal misalignment.

\begin{equation}
    \theta_{3D}^2 \sim  \int \theta_{PW}^2 ~ \dfrac{N_\gamma}{N_e} ~ \dfrac{\mathrm{d}N_b}{\mathrm{d} a_{0,\mathrm{eff}}} ~\mathrm{d}a_{0,\mathrm{eff}} ~/ \int \dfrac{N_\gamma}{N_e} ~ \dfrac{\mathrm{d}N_b}{\mathrm{d} a_{0,\mathrm{eff}}} ~\mathrm{d}a_{0,\mathrm{eff}}
\end{equation}

\subsection{Electron angular distribution}

As the photon divergence depends directly on the instantaneous electron divergence at emission time, estimating the final photon divergence can be accomplished with a simple integration, as shown in previous subsection. 
The asymptotic electron divergence is more challenging to estimate because of the accumulated diffusion of the distribution function due to the quantum stochasticity of photon emission. 
In \cite{vranicQuantumRadiationReaction2016}, a  scaling law was derived for the electron transverse momentum divergence in a PW as a function of average final energy $\gamma_f$ and energy spread $\sigma_f$, following the quantum radiation reaction on the emitting electrons. 
In \cite{vranicAreWeReady2019}, it was shown that if the precise values of final average energy and energy spread were used in the previously mentioned scaling law, it could accurately predict the final electron divergence. 
However, the estimate of the final divergence is sensitive to the accuracy of the estimates for $\sigma_f$ and $\gamma_f$.
The final electron divergence, measured as root-mean-squared angle in the electron momentum \cite{vranicQuantumRadiationReaction2016}, can be written as

\begin{equation}
    \theta_e^{PW} \sim \sqrt{\dfrac{2}{\pi}} \dfrac{a_0~\sigma_f}{\gamma_f^2}
\end{equation}

\noindent where the electron energy spread is upper-bounded by

\begin{equation}
    \sigma_{\mathrm{f}}^2 \leq 1.455 \times 10^{-4} \sqrt{I_{22}} ~\frac{\gamma_0^3}{\left(1+6.12 \times 10^{-5} ~\gamma_0 ~I_{22} ~\tau_0[\mathrm{fs}]\right)^3}.
    \label{eq:th_electrons_pw}
\end{equation}

\noindent Here, $I_{22}=10^{-22} ~I[\mathrm{W/cm}^{2}]$ is the laser intensity and $\gamma_f$ is the final electron energy given by equation \eqref{eq:PWCRR}. Similarly to the photon divergence, the equivalent electron divergence in 3D will be the root-mean-squared plane wave divergence weighted by the $a_{0,\mathrm{eff}}$ distribution.

\begin{figure}[H]
    \centering
        \includegraphics[width=360pt]{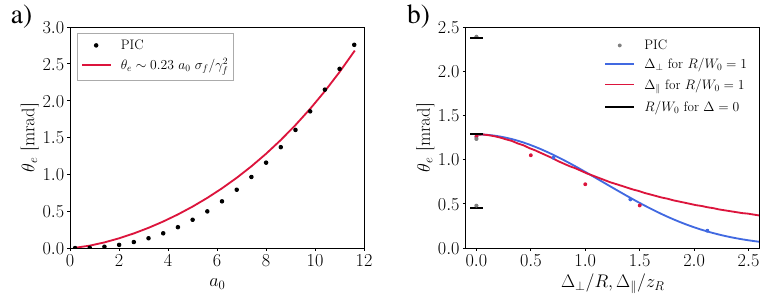}
        \caption{Left: electron angle rms in plane wave. Right: electron angle rms in focused laser setups.}
    \label{fig:thetaElectrons}
\end{figure}

In figure \ref{fig:thetaElectrons}, we compare this scaling law against PIC simulations, first in the case of a plane wave for different $a_0$ values, and then for focused Gaussian laser scattering with different collision offsets.

Equation \eqref{eq:th_electrons_pw} was originally derived for scattering against a laser pulse with a long, flat temporal envelope, where sufficient time would have passed for an equilibrium between radiative cooling and stochastic diffusion to have occurred. For the laser pulse durations considered in this work ($\ll 100 ~\mathrm{fs}$), this is not always satisfied. Consequently, we correct the scaling law with a multiplicative pre-factor to match the plane wave results with a shorter interaction time. 
The results are presented in figure \ref{fig:thetaElectrons}, where there is an overall good correspondence between the model and simulations, and once again we can incorporate the spatio-temporal misalignment. 

Knowing the asymptotic divergence of the electron beam can be very useful in experiments. 
For example, in \cite{tamburiniOnshotDiagnosticElectron2020, huQuantumstochasticityinducedAsymmetryAngular2020}, the authors explore the possibility of using the induced asymmetry in the electron angular distributions in a linearly polarized laser as a robust sign of quantum stochasticity broadening.
It may also be used to estimate the peak laser intensity during scattering, which cannot be directly measured.

\subsection{Positron yield in electron-laser scattering}

In electron-laser scattering for pair production, both nonlinear Compton Scattering and nonlinear Breit-Wheeler pair production have to be taken into account. These two stochastic processes have to occur in succession and within the crossing time of the laser pulse. In the \textit{soft-shower} regime (only one generation of pairs produced), the electrons emit photons of broad spectra and at different phases of the laser. As a consequence, photons are distributed not only in energy but also in their integrated probability of decaying into pairs. This makes theoretical modeling of these events somewhat more challenging than direct photon-laser scattering. 

An approximate scaling law for the positron yield (and average energy) in the scattering of electrons with a pulsed plane wave was derived in \cite{blackburnScalingLawsPositron2017}. This expression can then be implemented numerically into this framework (as we have done in \cite{amaroOptimalLaserFocusing2021}) to estimate the  positron yield in focused laser setups. 

\section{Conclusions}\label{sc:conclusions}

% findings intro
We present a semi-analytical model which offers a fast and computationally inexpensive method of predicting electron yield, photon yield, spectra and angular divergence in realistic electron scattering setups. 

% summary
We introduce the model by considering the interaction between a distribution of electrons   interacting with a focused intense laser pulse, assuming the probe particles perform ballistic trajectories. We then derive the closed-form electron spectra for different collision geometries after experiencing energy lost due to classical radiation reaction. 
Furthermore, we have shown that data from quasi-1d simulations of scattering against a plane wave with a temporal envelope can be recombined to yield the equivalent particle spectra in a focused laser setup, both for classical and quantum radiation reaction models.
In both cases, the simulations and the theory agree very well, demonstrating that the presented model can be used as a cheap and fast alternative to 3D-PIC simulations. 
By employing scaling laws for the final photon and electron divergence in a plane wave, we have also shown that these observables scenarios can also be predicted.
Besides Gaussian profiles, the model could also be extended to other beam profiles, such as Laguerre or Hermite-Gauss modes.

% incoherent radiation
The need for fast computation of observables in high-repetition-rate facilities requires a robust and efficient framework. The incoherent nature of the nonlinear Compton Scattering gamma radiation (independent between different electrons, but not within a single electron trajectory) allows calculation of the final spectrum as a sum of single-particle spectra (contrary to Thomson scattering where there will be interference patterns).
Having a sufficiently accurate forward model could enable the reconstruction of experimental parameters and profiles (beam and laser), and quantify the associated uncertainties. This will be particularly important in high repetition-rate facilities, where data needs to be processed quickly and can also be statistically relevant to train automated models. In the future, we will investigate this possibility.

% DLA
Another possible extension of this framework is the positron production at 90º of incidence for Direct Laser Acceleration (DLA). Contrary to a head-on collision, here the energy cutoff of the electron-positron pairs is no longer limited to the initial energy of the interacting electrons. If positrons are generated at low energies, they can be trapped, and accelerated along the laser propagation direction \cite{vranicMultiGeVElectronpositronBeam2018}.

% contributions
We strongly believe these findings will contribute to further analytical calculations on classical and quantum radiation reaction and pair production and help design future experiments to validate current models of Strong-Field QED.

\section{Acknowledgments}\label{sc:acknowledgments}

This article comprises part of the PhD thesis work of Óscar Amaro, which will be submitted to Instituto Superior Técnico, University of Lisbon.

The authors thank Mr. B. Barbosa, Dr. B. Martinez,  and Dr. L. I. I\~nigo Gamiz for fruitful discussions and proofreading the manuscript. This work was supported by the European Research Council (ERC-2015-AdG Grant No. 695088) and Portuguese Science Foundation (FCT) Grants No. CEECIND/01906/2018, PTDC/FIS-PLA/3800/2021, and UI/BD/153735/2022. Simulations were performed at the IST cluster (Lisbon, Portugal).

\vspace{10pt}
\noindent \textbf{References}
\bibliography{bibtex, bibtex_facilities}

\setcounter{equation}{0}
\renewcommand{\theequation}{A.\arabic{equation}}

\section*{Appendix A: PIC simulation parameters}\label{app:A}

\texttt{QScatter} (quick-scattering-toolkit) is an open-source library written in Python. Examples showing its usage are provided on GitHub at: \url{https://github.com/OsAmaro/QScatter}

In this appendix we specify the parameters of the simulations referred in this work. Unless specified, common to all simulations are the electron energy $\gamma_0=5000$, a laser wavelength of $\lambda$ = 0.8 micron ($\omega_0 = 2.35\times 10^{15}~s^{-1}$), linear polarization, a temporal envelope of the type $\sin^2$, and a corresponding rise time of the pulse of $50 ~1/\omega_p$. Additionally, when applicable, the laser spotsize was  $18.85~c/\omega_p$ and focused at the center of the simulation box. All macro-electrons and photons from the simulations are included in the particle distributions shown in the different plots of this work.

In section \ref{ssc:crr}, we consider setups where the classical model of radiation reaction is applied. For both the Wide and Thin beam geometries, the quasi-3d simulation box dimensions were $1100~c/\omega_p$ and $100~c/\omega_p$ with $5500$ and $500$ grid cells in the longitudinal and transverse directions, respectively. The time step was \texttt{dt} =  $0.02~1/\omega_p$ and particle information was retrieved at \texttt{tmax} =  $320~1/\omega_p$ (after interaction with the laser). The electron beam profile was flat top, with dimensions $356~c/\omega_p$ and $80~c/\omega_p$ in the Wide beam geometry, and with 20 particles-per-cell (ppc). For the Thin beam, the transverse dimension was much smaller than the laser spot size. In the case of the Short beam, the quasi-3d box dimensions were $300~c/\omega_p$ and $100~c/\omega_p$, with grid cells $1500$ and $500$, with a time step of \texttt{dt} = 0.02 $1/\omega_p$, and a total duration of \texttt{tmax} = 120 $1/\omega_p$. The electron beam profile was flat top longitudinally, with length 0.8 $c/\omega_p$, and Gaussian in the transverse direction, with radiae $13.33~c/\omega_p$ and $6.67~c/\omega_p$ for the different $R/W_0$ ratios, and \texttt{ppc}=320. 

In section \ref{ssc:qrr} we consider the quantum model of radiation reaction. For the quasi-1d simulations, the box size was $200~c/\omega_p$ with 2500 grid cells, a time step of \texttt{dt} =  $0.021~1/\omega_p$, and a total duration of \texttt{tmax} =  $130~1/\omega_p$. The electron density was flat top with length $4~c/\omega_p$, and \texttt{ppc}=$4000$. In the case of the synchronized Short Beam setup, the quasi-3d simulation box and temporal parameters were the same as in the Short beam with classical radiation reaction. The electron density profile was Gaussian in the transverse direction, radiae $13.33$, $26.66$ and $39.99~c/\omega_p$ respectively, and \texttt{ppc}=320. For the Short beam with parallel offset, the quasi-3d simulation box dimensions were $1100~c/\omega_p$ and $100~c/\omega_p$ with $5500$ and $500$ grid cells in the longitudinal and transverse directions, respectively. The time step was \texttt{dt} = $0.02~1/\omega_p$ and particle information was retrieved at \texttt{tmax} = $120~1/\omega_p$. The electron Gaussian density profile had a radius of $13.33~c/\omega_p$, with the initial center of the beam varying in the longitudinal direction in multiples of 89 $c/\omega_p$ for different offsets. In the case of the perpendicular offset, the simulation box had dimensions $300~c/\omega_p$ in the longitudinal direction and $240~c/\omega_p$ in both transverse directions, with 3000 and 1200 grid cells, respectively. Temporal resolution was \texttt{dt} = 0.02 $1/\omega_p$, with \texttt{tmax} = 120 $1/\omega_p$. The electron density profile was flat top in the longitudinal direction with length $0.8~c/\omega_p$, and Gaussian in the transverse direction with radius $13.33~c/\omega_p$. The initial center of the electron beam was varied along the \texttt{x2} direction in multiples of 13.33 $c/\omega_p$ for different offsets, and \texttt{ppc}=8.

\end{document}